# Slater Insulator in Iridate Perovskites with Strong Spin-Orbit Coupling

Q. Cui,[1] J.-G. Cheng,[1,2*] W. Fan,[3] A. E. Taylor,[4] S. Calder,[4] M.A. McGuire,[5] J.-Q. Yan,[5,6] D. Meyers,[7] X. Li,[2] Y. Q. Cai,[1] Y. Y. Jiao,[1] Y. Choi,[8] D. Haskel,[8] H. Gotou,[9] Y. Uwatoko,[9] J. Chakhalian,[10] A. D. Christianson,[4,11] S. Yunoki,[3,12,13] , J. B. Goodenough,[2] and J.-S. Zhou[2*]

[1] *Beijing National Laboratory for Condensed Matter Physics and Institute of Physics, Chinese Academy of Sciences, Beijing 100190, China*
[2] *Materials Science and Engineering Program, University of Texas at Austin, Austin, Texas 78712, USA*
[3] *Computational Condensed Matter Physical Laboratory, RIKEN, Wako, Saitama 351-0198, Japan*
[4] *Quantum Condensed Matter Division, Oak Ridge National Laboratory, Tennessee 37831, USA*
[5] *Materials Science and Technology Division, Oak Ridge National Laboratory, Tennessee 37831, USA*
[6] *Department of Materials Science and Engineering, University of Tennessee, Knoxville, Tennessee 37996, USA*
[7] *Department of Physics, University of Arkansas, Fayetteville, AR 72701, USA*
[8] *Advanced Photo Source, Argonne National Laboratory, Argonne, Illinois 60439, USA*
[9] *Institute for Solid State Physics, University of Tokyo, 5-1-5 Kashiwanoha, Chiba 277-8581, Japan*
[10] *Department of Physics and Astronomy, Rutgers University, 136 Frelinghuysen Road, Piscataway, New Jersey 08854, USA*
[11] *Department of Physics and Astronomy, University of Tennessee, Knoxville, Tennessee 37966, USA*
[12] *Computational Materials Science Research Team, RIKEN Advanced Institute for Computational Science (AICS), Kobe, Hyogo 650-0047, Japan*
[13] *Computational Quantum Matter Research Team, RIKEN Center for Emergent Matter Science (CEMS), Wako, Saitama 351-0198, Japan*

jgcheng@iphy.ac.cn and jszhou@mail.utexas.edu

# Abstract

The perovskite $SrIrO_3$ is an exotic narrow-band metal owing to a confluence of the strengths of the spin-orbit coupling (SOC) and the electron-electron correlations. It has been proposed that topological and magnetic insulating phases can be achieved by tuning the SOC, Hubbard interactions, and/or lattice symmetry. Here, we report that the substitution of nonmagnetic, isovalent $Sn^{4+}$ for $Ir^{4+}$ in the $SrIr_{1-x}Sn_xO_3$ perovskites synthesized under high pressure leads to a metal-insulator transition to an antiferromagnetic (AF) phase at $T_N \geq 225$ K. The continuous change of the cell volume as detected by x-ray diffraction and the λ-shape transition of the specific heat on cooling through $T_N$ demonstrate that the metal-insulator transition is of second-order. Neutron powder diffraction results indicate that the Sn substitution enlarges an octahedral-site distortion that reduces the SOC relative to the spin-spin exchange interaction and results in the type-G AF spin ordering below $T_N$. Measurement of high-temperature magnetic susceptibility shows the evolution of magnetic coupling in the paramagnetic phase typical of weak itinerant-electron magnetism in the Sn-substituted samples. A reduced structural symmetry in the magnetically ordered phase leads to an electron gap opening at the Brillouin zone boundary below $T_N$ in the same way as proposed by Slater.

Magnetism as described by the Heisenberg Hamiltonian originates from interatomic electron-electron correlations (EEC). The correlations can also facilitate a metal-insulator transition, *i.e.* a Mott transition.[1, 2] Alternatively, a reduced translation symmetry due to antiferromagnetic ordering in a metal with a half-filled band can also lead to a metal-insulator transition as proposed by Slater.[3] Whether the Mott physics can actually cover the second case remains controversial. The case of a Slater insulator is extremely rare; $Cd_2Os_2O_7$ and $NaOsO_3$ have recently been actively discussed in the context of a Slater insulator.[4-6] Issues such as whether the "all-in and all-out" spin ordering would lower the crystal symmetry in $Cd_2Os_2O_7$ and whether a relatively sharp change of lattice parameters at $T_N$ would reflect the effect of EEC on the transition in $NaOsO_3$ have made it difficult to distinguish a Mott versus a Slater transition in these experiments. The metal-insulator transition in $SrIr_{1-x}Sn_xO_3$ reported in this Letter brings a fresh case to the dialog.

The 4d and 5d transition-metal oxides (TMO) bring in a strong intraatomic spin-orbit coupling (SOC) on top of the dominant interatomic spin-spin interactions seen in 3d TMO.[7, 8] Rich physical phenomena in the 4d and 5d TMO have drawn great attention in the physics community.[8-16] In the Ruddlesden-Popper (RP) series $Sr_{n+1}Ir_nO_{3n+1}$ (n=1, 2, ∞) consisting of rock-salt layers of SrO and PV layers of $SrIrO_3$, the electron bandwidth can be tuned by the structural dimensionality from a Pauli paramagnetic metal in the PV $SrIrO_3$ (n=∞) [17, 18] to insulators with layered structures (n = 1, 2) showing a magnetic transition.[19-21] The resistivity shows the activated behavior in both the paramagnetic and magnetic phases in these layered iridates.[19, 20, 22, 23] Since a strong SOC in 5d TMO mixes up $t_2$ orbitals with spin states to form a filled $J_{eff} = 3/2$ band and a half-filled $J_{eff} = 1/2$ band in $Sr_2IrO_4$, the enhanced EEC in this layered iridate splits the half-filled $J_{eff} = 1/2$ band to form a Mott insulator.[9, 10] Dimensional bandwidth control [21] has been implemented in the superlattice structures between two PV blocks of $SrTiO_3$ and $SrIrO_3$ by epitaxial growth.[24] In the PV structure of $AMO_3$, the 180° M-O-M bonds in the 3D framework of corner-shared octahedra provide the electronic transport. The bandwidth can be tuned by either bending the M-O-M bond from 180° or truncating M-O-M array along the *c* axis as in the RP series. Recent ARPES studies on PV $SrIrO_3$ films have shown an unusually narrow bandwidth near the Fermi level due to a confluence of SOC and the octahedral rotations.[25] Several distinct topological and magnetic insulating phases have also been proposed in PV

SrIrO$_3$ by tuning the SOC, EEC, and/or lattice symmetry.[26, 27] In this work, we fine tune the electron bandwidth in the PV SrIrO$_3$ by introducing diamagnetic Sn$^{4+}$ randomly into the 3D Ir-O-Ir array. For compositions of x ≥ 0.1 in SrIr$_{1-x}$Sn$_x$O$_3$, the PV phase exhibits a remarkable metal-insulator (MI) transition to an antiferromagnetically ordered phase at T$_N$.

SrIrO$_3$ crystallizes in the orthorhombic perovskite structure with the space group *Pbnm* if it is synthesized under high pressure.[28] SrSnO$_3$ can be prepared with the *Pbnm* structure, which makes it possible to have a continuous solid solution between SrIrO$_3$ and SrSnO$_3$. All PV SrIr$_{1-x}$Sn$_x$O$_3$ (0 ≤ x ≤ 0.5) samples in this study were prepared at 6 GPa and 1000 °C. The structure, specific heat and magnetization of each composition were characterized; detailed information can be found in the Supplemental Material (SM) [url],[29] which includes Refs.[30-32] The Ir$^{4+}$ oxidation state in all samples has been confirmed with Ir-L$_3$ edge X-ray Absorption Spectroscopy (XAS), Fig. S1 in SM.[29]

Fig.1 displays how the transport and magnetic properties of SrIr$_{1-x}$Sn$_x$O$_3$ change with Sn substitution. The parent compound SrIrO$_3$ is a Pauli paramagnetic metal as is seen from the resistivity and magnetic susceptibility, which are consistent with data in the literature.[17, 18] However, the temperature dependence of thermoelectric power does not follow the Mott diffusive formula for a metal. An enhancement of |S| that peaks at ~175 K cannot be rationalized by the phonon-drag effect, which should occur at a small fraction of the Debye temperature.[33] The resistivity shows an anomaly at approximately the same temperature. In metallic PV SrRhO$_3$ where the SOC effect is relatively weak, an enhancement near 180 K can still be discerned in S(T), but the enhancement is much smaller than that found in SrIrO$_3$. This comparison highlights the possible SOC effect on the thermoelectric power, which deserves a further study. Introducing diamagnetic Sn$^{4+}$ in the SrIr$_{0.9}$Sn$_{0.1}$O$_3$ sample of Fig.1(b) does not destroy the metallic phase completely, but it induces a metal-insulator transition at T ≈ 225 K; the resistivity increases by more than 4 orders of magnitude as the sample is cooled down to 4 K from T$_N$. Correspondingly, a transition to a weak ferromagnetic phase can be discerned from the magnetization measurement and confirmed by neutron diffraction at the same temperature. The thermoelectric power in the paramagnetic phase of the x = 0.1 sample is positive and initially decreases slightly as temperature decreases to T$_N$, resembling the behavior of a normal metal. On cooling though T$_N$, the slope of S(T) changes, which is followed by a sharp increase at T < T$_N$. The metal-insulator transition moves to higher temperatures progressively as the Sn concentration increases, as can be seen in samples x = 0.2 and 0.3. For samples with x ≥ 0.2, however, the resistivity in the paramagnetic phase becomes activated and charge carriers are trapped out as shown in the S(T) as temperature decreases; the high concentration of $Sn^{4+}$ appears to bring in a higher degree of disorder to scatter charge carriers. Nevertheless, the antiferromagnetic transition still causes a clear anomaly in both S(T) and ρ(T) at $T_N$, which is in sharp contrast to what is found in the $J_{eff}= 1/2$ Mott insulator $Sr_2IrO_4$.

The central issue of this study is to identify the driving force for the metal-insulator transition. A Mott transition is due to inter-site EEC that introduce a gap in the single-electron band structure of a single-valent compound.[1] The transition is from an enhanced Pauli paramagnetic metal to a Curie-Weiss insulator and is of first-order if orbital degeneracy is present. Moreover, $T_{IM}$ is in most cases higher than or equal to a magnetic transition temperature. For the case of $Sr_2IrO_4$ where strong SOC creates an occupied $J_{eff}= 3/2$ band and a half-filled $J_{eff}= 1/2$ band,[9, 10] the EEC is responsible for splitting the half-filled $J_{eff}= 1/2$ band although the $T_{IM}$ above room temperature remains to be identified. Under the treatment of Hartree-Fock approximation, an electron moves in the potential built by other electrons. Without the EEC, an electron is dressed with the spin-spin correlation that modifies the potential in the crystal lattice if the crystal becomes antiferromagnetically ordered. Slater[3] has proposed that the change of translational symmetry in an antiferromagnetic phase opens up a gap at a Fermi energy pinned at the Brillouin-zone boundary if it has a half-filled band. The transition is of second-order and an antiferromagnetic interaction in the paramagnetic phase is necessary to lead to the spin ordering.

Back to the case of $SrIr_{1-x}Sn_xO_3$, the origin of the magnetism in the Sn-substituted samples is rooted in the parent oxide $SrIrO_3$ although it is a paramagnetic metal. The EEC effect can be probed through the ratio of $\gamma/\gamma_0$, where γ is the Sommerfeld coefficient of the specific heat (see Fig. S3 and Table S1 for details) and $\gamma_0$ is the calculated electronic contribution from the band structure.[21] For $SrIrO_3$, a ratio $\gamma/\gamma_0 \approx 1.1$ indicates a modest correlation enhancement. However, the ratio of the temperature-independent term $\chi_0$ obtained from the magnetic susceptibility of Fig.1(a) versus the Pauli magnetic susceptibility $\chi_P$ calculated from the band structure, $\chi_0/\chi_P = 5.7$, shows a much stronger Stoner enhancement as in $LaNiO_3$,[34] which places $SrIrO_3$ on the verge of a Stoner instability. The unusual increase of magnetization below 50 K shown in Fig.1(a) may signal such an instability.

Determining how the $Sn^{4+}$ substitution in $SrIr_{1-x}Sn_xO_3$ further enhances the interatomic magnetic coupling so as to eventually trigger the magnetic transition relies on our understanding of (1) the substitution effect on the crystal structure and (2) the explicit role of SOC on the spin-spin exchange interaction. Depending on the strength of SOC, the electronic state can be governed by either the crystal field splitting or the strong SOC effect that makes J a good quantum number. In a picture of the former, relatively weak SOC effect only mixes up the xy, yz, zx orbitals and spin states into two groups with effective J= ½ and 3/2 separated by $\lambda \mathbf{L} \cdot \mathbf{S}$. [9, 10, 29] The low-spin $d^5$ electron configuration on $Ir^{4+}$ places one hole in the two-fold degenerate $J_{eff}=1/2$ state. The SOC strength, however, is sensitive to local structural distortion. The intrinsic distortion of the *Pbnm* $AMO_3$ structure splits the M-O bonds of an $MO_6$ octahedron into long, medium, and short bonds.[35] Refinements of neutron powder diffraction[29] show that the Ir-O bond lengths in the *ab* plane come closer to one another and the bond length ratio is $l_c/l_{ab} < 1$ in $SrIrO_3$;[18] which is still compatible to preserve L in $t_{2g}$ orbitals. The important observation in Fig.2(a) is that the bond length splitting progressively increases with increasing Sn concentration. Specifically, the bond length splitting in the *ab* plane enlarges in the Sn-substituted samples; therefore the orbital angular momentum L found in $SrIrO_3$ may not be preserved in zero order in the Sn substituted samples. Although the bond length splits in the *ab* planes, $l_c/l_{ab} < 1$ for an averaged $l_{ab}$ remains. One may say that the bond-length splitting in the x = 0.2 sample could favor the orbital with **L** pointing to $O_{21}$. As explained in the following, an easy axis along the Ir-$O_{21}$ bonding direction is not allowed in the *Pbnm* crystal structure. As pointed out by Kanamori [36] the effects of SOC include (a) the magnetic susceptibility no longer follows the Curie-Weiss law; (b) the SOC competes with the spin-spin exchange interaction so as to lower the Néel temperature. The evolution of local structural distortion in $SrIr_{1-x}Sn_xO_3$ reduces the orbital moment L so as to lower the competition effect of SOC to the spin-spin exchange interaction. Moreover, the band narrowing effect due to the random occupation of diamagnetic $Sn^{4+}$ at $Ir^{4+}$ sites enhances EEC and therefore the interatomic exchange interactions.

For the x = 0.1 sample, the paramagnetic magnetic susceptibility is enhanced in comparison to that of SrIrO$_3$ and, most importantly, it becomes temperature dependent although the SOC effect is still strong enough to render an unusually large Weiss constant if fit to the Curie-Weiss (CW) law in Fig.2(b). The trend for $\chi^{-1}(T)$ to become more temperature dependent is clearer as x increases further. For the x = 0.5 sample, a CW fitting gives a $\mu_{eff}$ = 1.53 $\mu_B$ per Ir, which is close to the spin-only value $\mu_{eff}$ = 1.73 $\mu_B$ per Ir. Moreover, the Weiss constant $|\theta|$ = 602 K obtained by fitting $\chi(T)$ to a CW law is much closer to the Néel temperature than that for samples with x < 0.5.[29] The smaller magnitude of $\chi(T)$ in these heavily substituted samples is due to dilution with diamagnetic $Sn^{4+}$ (all B-site ions are counted in the calculation of the magnetization per mole). These observations indicate that the magnetic spin-spin interaction in the paramagnetic phase increases due to a combination of an enhanced exchange interaction and the reduction of the competing effect from the SOC in SrIr$_{1-x}$Sn$_x$O$_3$ with more distorted crystal structure.

The temperature dependences of the cell volume V(T) and lattice parameters of x = 0.1 and 0.2 samples shows a second order transition with smooth changes on crossing $T_N$, as seen in Fig.3(a,b), which rules out the possibility of a Mott transition. Although V(T) of NaOsO$_3$ changes smoothly on crossing $T_N$, the lattice parameters show a dramatic change at $T_N$ in addition to more abrupt changes of $\rho(T)$ and M(T),[6] which are in sharp contrast to that in SrIr$_{1-x}$Sn$_x$O$_3$. The specific heat measurement also indicates a second order transition at $T_N$ in SrIr$_{1-x}$Sn$_x$O$_3$.[29] After identifying the magnetic coupling in the paramagnetic phase and a second order transition, we come to the question whether the magnetic transition indeed lowers the crystal symmetry as required for a Slater insulator.[3] Fig. 3(c) displays results of neutron diffraction made on the x = 0.2 sample at 100 K to 300 K. Based on the orthorhombic *Pbnm* structure, the peak at $Q$ = 1.377(5) Å$^{-1}$ is from (0 1 1) and (1 0 1) reflections. (011) is a forbidden structural reflection; the intensity of (101) reflection is in the same level as the background and shows a negligible dependence on temperature as seen from the structural refinements. A peak centered at Q = 1.377(5) Å$^{-1}$ was observed at 100 K but is absent in the data at 300 K. The evolution of this superlattice peak as a function of temperature in Fig. 3(d) tracks the M(T) behavior in Fig. 1(c) nicely. These observations indicate that the signal centered at $Q$ = 1.377(5) Å$^{-1}$ is magnetic in origin. Therefore, the magnetic transition lowers the translational symmetry by enlarging the unit cell. Since the reflections center at $Q$ = 1.377(5) Å$^{-1}$ are very weak, we cannot determine the moment orientation based on the present powder measurement. However, a symmetry analysis shows that a peak at (0 1 1) and/or (1 0 1) is only allowed by two products ($G_xF_z$) and ($F_xG_z$) in two irreducible representations of G-type AF ordering in the *Pbnm* structure,[37] as shown by Fig.S7 in SM together with the LDA+U+SOC calculations which also verify the ground state with G-type AF ordering. In the AF structure of ($G_xF_z$), the easy axis is pointing to the middle of the connection between $O_{21}$ and $O_{22}$, which has been seen in 3d TMO, for example $LaCrO_3$.[38] In these cases, the easy axis is determined by the global symmetry instead of the local site distortion. In the 5d oxides, however, the spin direction is dictated by the orbital moment direction which is coupled strongly to the octahedral distortion.[39] For example, the easy axis of spin on $Ir^{4+}$ in the post-Pv $CaIrO_3$ is pointing to the same corner-shared oxygen as an orbital moment L which is preserved by the crystal symmetry.[40, 41] In the PV structure of $SrIr_{1-x}Sn_xO_3$, the only direction of L that can be accommodated by the local structural distortion is along the *c* axis, so it is the easy axis for spins. Therefore, the ($F_xG_z$) canting structure is preferred.

The evolution of SOC in the AF insulator phase as a function of x (≥0.1) has been probed by XAS/XMCD of Fig.4(a). The branching ratio (BR) from XAS, defined here as $I_{L3}/I_{L2}$, gives important information regarding the expectation value of ⟨L·S⟩; a BR value larger than the statistical value of 2 indicates a strong spin-orbit coupling. For samples with x = 0.1, 0.2, 0.3, 0.5, the BR is found to be between 4.8 and 5.3 ± 0.21, indicating a strong SOC effect as also seen in several octahedral coordinated iridates of Table S2.[42-44] By using ⟨L·S⟩ = ⟨$n_h$⟩ (BR – 2)/ (BR +1), where ⟨$n_h$⟩ is the averaged number of holes, we obtain an expectation value of ⟨L·S⟩ from 2.6± 0.15 to 3.0 ± 0.15. The departure from the expected value of 1 for a $J_{eff}$ = 1/2 system was previously shown to indicate the SOC is strong enough to mix some $e_g$ orbitals into the $t_{2g}$ states.[42] Since all measurements were performed at 150 K, ⟨L·S⟩ versus x at 150 K will cross the metal-insulator phase boundary in the phase diagram of Fig.4(b). A clear jump of ⟨L·S⟩ on crossing the phase boundary indicates that SOC is enhanced in the insulator phase, which is consistent with the argument by Goodenough and Kanamori[36, 45] and justifies that we have treated the paramagnetic phase above based on the crystal-field dominated picture. In the AF insulator phase with a more or less constant $T_N$, ⟨L·S⟩ reduces, which reflects the effect of structural distortion on SOC. Because of an abrupt increase of SOC on crossing MIT, the $J_{eff}$=1/2 state is reinforced so as to make the Sn substituted $SrIr_{1-x}Sn_xO_3$ a $J_{eff}$=1/2 Slater insulator.

Given a magnetic moment ~ 1 $\mu_B$/Ir from fitting the $\chi(T)$ to a CW law in x=0.5 sample, an extremely small saturation moment (about 0.043 $\mu_B$/Ir) extracted from the magnetization curve at 150 K of Fig.S5 corresponds to a small canting angle $\alpha \approx 2.5°$ at $Ir^{4+}$ sites in the AF spin structure of ($F_xG_z$) shown as the inset of Fig. 2(a). The spin canting angle is far smaller than the octahedral rotation angle $\phi \approx 20°$ from our structural study of these iridate perovskites; a $\phi/\alpha \sim 0.1$ is obtained. In contrast, Jackeli and Khaliullin have shown that the ratio $\phi/\alpha$ varies in a small range 0.75 to 2 depending on the ratio of *c/a* of a tetragonally distorted octahedron in $Sr_2IrO_4$.[39] Because of stronger SOC in the $J_{eff}$=1/2 Mott insulator $Sr_2IrO_4$ than that in $SrIr_{1-x}Sn_xO_3$ as derived from the XMCD results, see Table S2, the octahedral rotation clearly dictates the spin canting in $Sr_2IrO_4$. A distinct difference of $\phi/\alpha$ could be another factor to distinguish a $J_{eff}$ = 1/2 Mott insulator from a $J_{eff}$ = 1/2 Slater insulator where the SOC effect is relatively weak in the band state.

In conclusion, the interplay of electron-electron correlations and a strong SOC places the Pv $SrIrO_3$ phase on the verge of a Stoner instability. The Sn substitution strengthens the spin-spin exchange interaction while reducing the competing SOC effect by making $IrO_6$ octahedra more distorted, which eventually leads to a temperature-driven metal-insulator transition to a type-G antiferromagnet at $T_N$. A smooth change of the cell volume on crossing $T_N$ rules out the possibility of a Mott transition in the iridate system with orbital degeneracy. An enhanced SOC on crossing the transition from paramagnetic metal to the magnetic insulator phase indicated by the XAS and XMCD results support the picture that the $t_{2g}$ orbitals and spin states are regrouped into $J_{eff}$ = 3/2 band and $J_{eff}$ = ½ states. A gap opening in a half-filled $J_{eff}$ = 1/2 band with the orbital degeneracy for low spin $Ir^{4+}$ with five d electrons due to the electron scattering at the Brillouin zone boundary by an enlarged periodicity of the potential in the antiferromagnetic phase provides a case that may be used to justify the mechanism proposed by Slater.

## Acknowledgments

This work was supported by the National Basic Research Program of China (Grants No. 2014CB921500), the National Science Foundation of China (Grants No. 11304371, 11574377), the Strategic Priority Research Program of the Chinese Academy of Sciences (Grant XDB07020100), and the Opening Project of Wuhan National High Magnetic Field Center (Grant

No.2015KF22), Huazhong University of Science and Technology. JSZ and JBG were supported by the NSF-DMR-1122603 and the Welch Foundation (F-1066). The research at Oak Ridge National Laboratory was supported by the U.S. Department of Energy, Office of Science, Basic Energy Sciences, Materials Science and Engineering Division (MAM and JQY) and Scientific User Facilities Division (AET, SC, and ADC). SY was supported by Grant-in-Aid for Science Research from MEXT Japan under the Grant No. 25287096.

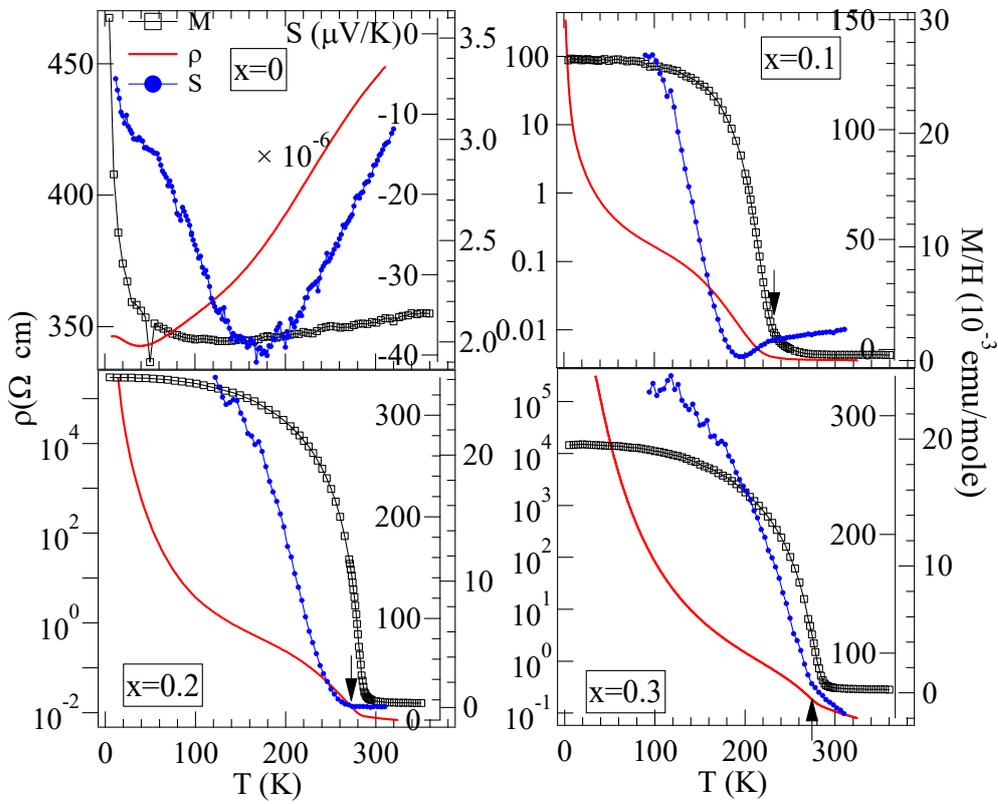

Fig. 1 Temperature dependence of the magnetization M/H, resistivity $\rho$, and the thermoelectric power S(T) for $SrIr_{1-x}Sn_xO_3$ ($0 \leq x \leq 0.5$) samples. Resistivity curves on cooling down and warming up overlap. The right label is for the magnetization and the inside label is for thermoelectric power. Arrows point to the transition temperature as detected by the change of thermoelectric power.

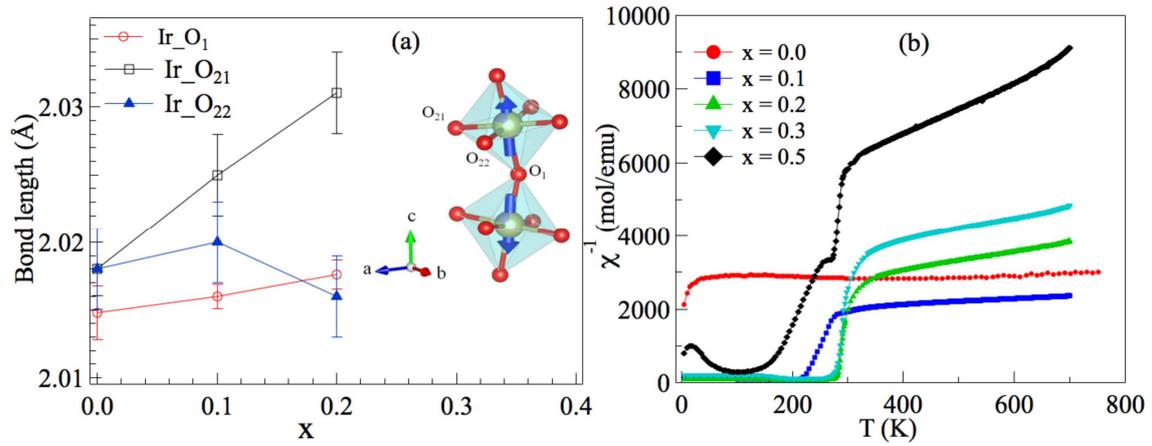

Fig.2 (a) The bond lengths in an Ir(Sn)$O_6$ octahedron in SrIr$_{1-x}$Sn$_x$O$_3$ resolved by neutron powder diffraction; the data for the x = 0 sample are after ref.14. The inset: schematic drawing of octahedra, their tilting configuration in the structure, and the relationship to the spin canting. (b) Temperature dependence of magnetic susceptibility for SrIr$_{1-x}$Sn$_x$O$_3$.

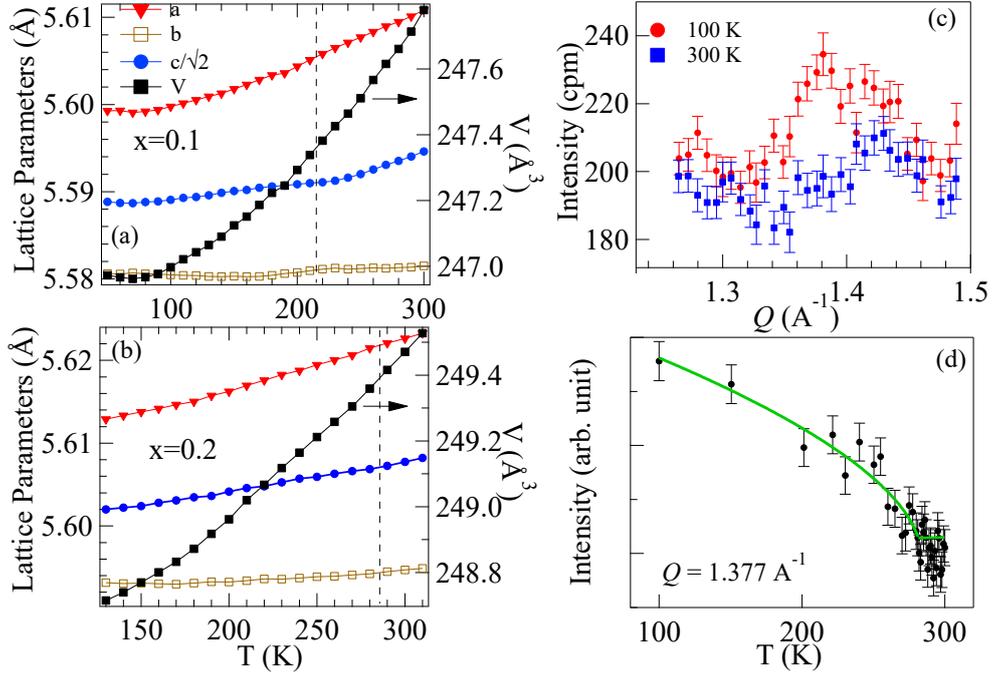

Fig. 3 (a,b) Temperature dependences of lattice parameters for x= 0.1 and 0.2 samples of SrIr$_{1-x}$Sn$_x$O$_3$ determined by X-ray diffraction; dashed lines indicate the MI transition temperatures; (c) Q dependence of neutron powder diffraction intensity measured at 100K (circles) and 300K (diamonds), normalized to counts per minute (cpm) for x=0.2 sample. (d) Temperature dependence of intensity measured at the fixed position $Q = 1.377$ Å$^{-1}$. The line is a guide to the eye.

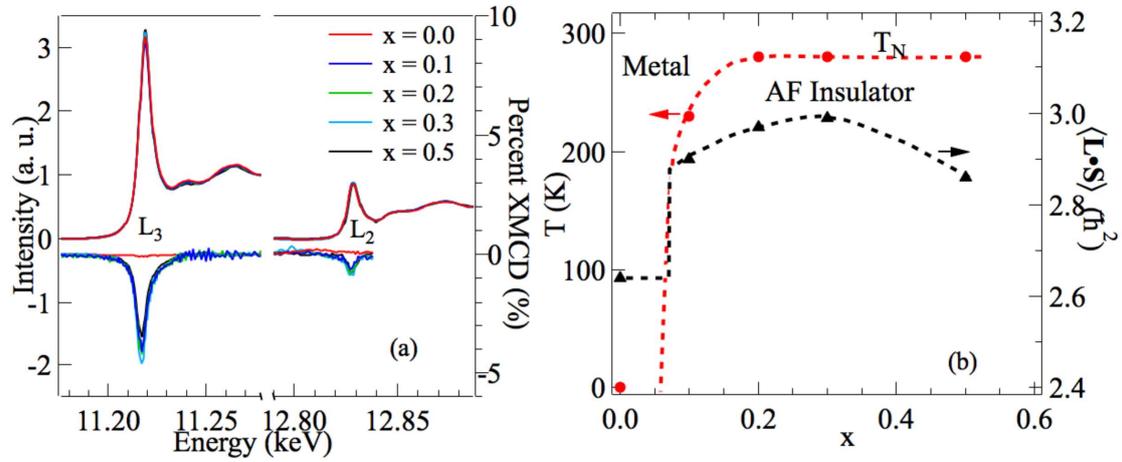

Fig. 4 (a) XAS and XMCD spectra of SrIr$_{1-x}$Sn$_x$O$_3$ ($0 \leq x \leq 0.5$) samples at 150 K, and (b) The phase diagram of SrIr$_{1-x}$Sn$_x$O$_3$ together with the expectation value of spin-orbit coupling.